# Real Time Electron Tunneling and Pulse Spectroscopy in Carbon Nanotube Quantum Dots

*Georg Gotz , Gary A. Steele, Willem – Jan Vos, Leo P. Kouwenhoven*

Kavli Institute of Nanoscience, Delft University of Technology, P.O. Box 5046, 2600 GA Delft, The Netherlands

G.T.J.Goetz@tudelft.nl

ABSTRACT: We investigate a Quantum Dot (QD) in a Carbon Nanotube (CNT) in the regime where the QD is nearly isolated from the leads. An aluminum single electron transistor (SET) serves as a charge detector for the QD. We precisely measure and tune the tunnel rates into the QD in the range between 1 kHz and 1 Hz, using both pulse spectroscopy and real – time charge detection and measure the excitation spectrum of the isolated QD.

A quantum dot (QD) defined in a carbon nanotube (CNT) is a very interesting and unique physical system for studying individual electron spins [1, 2, 3, 4, 5, 6]. In particular, the spin relaxation and coherence times are expected to be as long as seconds [1, 2], which makes this system attractive for quantum information processing. However, both precise control over the tunnel rate into a QD and real – time read out of the charge state of the QD have not been demonstrated yet for CNTs.

QDs can be defined in CNTs by using top gates (TGs) as shown in Figure 1. Suitable voltages applied to these TGs create local tunnel barriers in semiconducting CNTs. In this way, single and double QDs have been realized [7, 8, 9]. In this letter we use TGs to precisely tune the tunnel rates into a CNT–QD all the way down to ~1 Hz. We use a metallic SET as a charge detector, sensitive to single electron charges in the CNT–QD [10, 11, 12], since transport measurements are not possible at such low tunnel rates. Finally, we measure the excitation spectrum of a nearly isolated CNT–QD.

Samples are fabricated on highly p–doped Si substrates with 280 nm thermally grown silicon oxide, such that the Si substrate can serve as a global back gate (BG). CNTs are grown from Fe/Mo catalyst islands using chemical vapor deposition [13]. The CNTs are located with atomic force microscopy and CNTs with diameters $\leq 4$ nm (probably single walled) are chosen for further sample fabrication. First, ohmic contacts are made with thermally evaporated Pd (15 nm). The entire sample is then coated with 35 nm $Al_2O_3$, deposited with Atomic Layer Deposition. Room temperature conductivity measurements as function of applied BG voltage allow us to select semiconducting CNTs. We define two TGs and one SG by evaporating Ti/Au (10/20 nm). In close proximity ($< 500$ nm) to the CNT we fabricate an aluminum single electron transistor (SET) using a standard double angle evaporation technique [14]. Figure 1a shows an atomic force microscopy image of a device from the same fabrication run as the one from which we present data here. After evaporation of Ti/AuPd bonding pads, devices are wire bonded and cooled down in a dilution refrigerator with a base temperature of ~50 mK. The CNT–SG is connected to a bias tee (at room temperature), allowing us to apply DC– and AC–voltages at the same time to this gate. We show data from one device, but similar circuits have been realized three times indicating the reproducibility of our approach.

First, we characterize our device. Figure 1b shows the current through the CNT, $I_{CNT}$, versus back gate voltage, BG, and versus TG voltages. The two TGs show similar behavior, albeit with a smoother pinch-off curve for $TG_2$. The oscillations in the trace for $TG_1$ are likely due to disorder at low electron densities [8]. This residual disorder leads to resonances and charging effects in the small regions underneath the TGs.

Despite the remaining disorder, we create a QD in the CNT segment between the two TGs by choosing appropriate TG voltages (Figure 2a, inset). The addition energy of a few meV is typical for a CNT–QD with a length of ~500 nm, i.e. consistent with the distance between our TGs [7, 8, 9]. Moreover, we find that the QD couples equally to both TGs (measurements not shown here), which indicates that the QD is indeed formed between the two TGs. From the voltages applied to the BG and SG, we estimate the number of electrons in the CNT–QD, $N$, to be several 100.

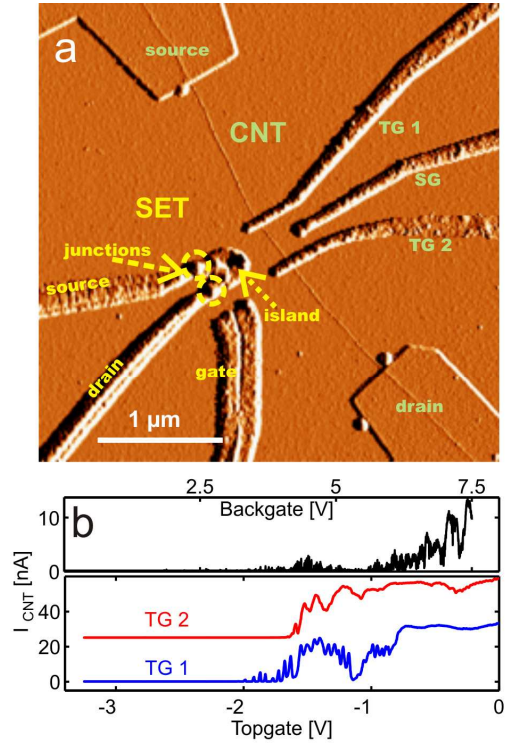

**Figure 1** Sample characterization **a)** Atomic force microscopy picture of a device similar to the one from which we present measurements. The semiconducting CNT is contacted with Pd ohmic contacts, separated by ~3 μm. Two 40 nm wide top gates ($TG_1$ and $TG_2$) cross the CNT with a 400 nm separation with a side gate (SG) located between them. On the lower side of the picture an Al–SET is fabricated close to the CNT segment between the two TGs. (In contrast to the measured device, this picture shows a sample with a dielectric only below the TGs instead of an $Al_2O_3$ layer covering the entire sample. Therefore, the CNT is clearly visible here.) **b) top:** Current through the CNT ($V_{SD}^{CNT} = 2\,mV$) as function of applied back gate voltage (all other gates at 0 V) **bottom:** Current through the CNT ($V_{SD}^{CNT} = 10\,mV$) as function of either the voltage applied to $TG_1$ (blue) or $TG_2$ (red). Traces are offset for clarity. The other TG is set to 0 V, the back gate to 7.5 V. Both TGs pinch off the current for voltages ≤ -2 V, whereas a large current flows for voltages ≥ -0.75 V. This demonstrates tunable, local tunnel barriers.

We use the SET as a charge detector for our CNT–QD [10, 11, 12]. The SET consists of a small aluminum island, connected to source and drain via $Al_2O_3$ tunnel junctions, with a charging energy of a few 100 µeV. An additional gate allows us to tune the electrical potential on the SET island. In all measurements presented here, a perpendicular magnetic field of at least 0.2 T was applied, in order to keep the aluminum SET in the normal conducting state. Figure 2a shows simultaneous measurements of the CNT–current, $I_{CNT}$, and the SET–current, $I_{SET}$, while changing the voltage on SG. The sudden shifts in $I_{SET}$ reflect a change of one electron charge on the CNT–QD. The change in SET–current at such a phase shift is maximal when the SET is at the steepest slope of its oscillation, meaning the sensitivity is maximal at these points. In all measurements, we apply an appropriate voltage to the SET side gate, such that the investigated electron transition of the CNT–QD coincides with the SET at a point of highest sensitivity. For this sample, the coupling between SET and CNT–QD is about 0.2 (i.e. adding one electron to the CNT–QD induces an effective charge of 0.2 electrons on the SET island). The low frequency sensitivity to electrons on the CNT–QD is about $10^{-2}$ $e/\sqrt{Hz}$, demonstrating a high quality of our charge detector. The sensitivity is obtained from noise measurements of the SET–current at the operation point and is limited by the charge noise in our system.

The charge sensing also works for opaque tunnel barriers. An extreme example is shown in Figure 2b with a real–time measurement [11, 15, 16] of the SET–current when the CNT–QD potential is tuned close to a charge transition. We observe single electrons tunneling in and out the CNT–QD in real time with tunneling rates around $\Gamma = 1$ Hz. The fact that we only see two stable levels in $I_{SET}$ over such long timescales demonstrates the excellent stability of the combined system of CNT–QD and charge detector. In particular, we do not observe any bistabilities or disturbing effects from nearby charge traps. We precisely control the tunneling rate by the voltages applied to the topgates (figure 2b, inset). This demonstrates that it is possible to tune the TG–controlled barriers to tunnel rates as slow as $\Gamma \sim$ Hz, i.e. the interesting range of the expected life time of electron spins.

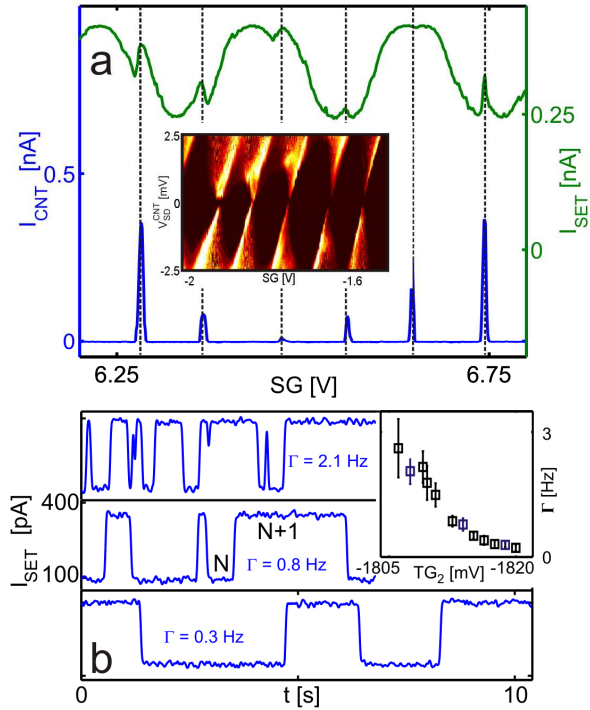

**Figure 2** Charge detection of single electrons in the CNT–QD **a)** Simultaneous measurement of the current through the CNT (blue, $V_{SD}^{CNT} = 0.5\,mV$) and SET (green, $V_{SD}^{SET} = 70\,\mu V$). The CNT–current shows Coulomb peaks when one more electron is added to the QD. The SET–current, $I_{SET}$, shows two features: first, we see periodic Coulomb oscillations of the current, resulting from the capacitive coupling of the SG to the SET island. Additionally, these oscillations undergo a number of sharp phase shifts. The phase shifts coincide with the Coulomb oscillations of the CNT–QD and are caused by the additional electron charge on the CNT–QD. If this phase shift happens at a point of steepest slope of the oscillations in $I_{SET}$, the change in $I_{SET}$ and thus the sensitivity is maximal. **Inset:** differential conductance *dI/dV* of the CNT–current as function of $V_{SD}^{CNT}$ and SG voltage. Regular Coulomb diamonds indicate a single QD with an addition energy of ~ 3meV, formed between the two TGs. **b)** Real–time detection of electron tunneling into the CNT–QD: Almost entirely pinched–off TGs result in extremely low tunneling rates, such that single tunneling events are visible in the SET–current. The two levels in the SET–current correspond to the two configurations of the CNT–QD being occupied with *N* or *N+1* electrons. The three traces are taken for different voltages, applied to TG$_2$. The other topgate is completely closed (TG$_1$ = -2400 mV) while we use TG$_2$ to precisely tune the tunneling rate into the QD to values around 1 Hz (top to bottom). **Inset:** Tunneling rate as a function of the voltage on TG$_2$. The rates are obtained from real-time traces. The blue data points correspond to the three traces shown in b).

Real–time detection of electron tunneling is a direct and elegant way for measuring tunneling rates of nearly closed QDs. However, it is time consuming in terms of data analysis and technically demanding already for frequencies in the kHz range or higher. A more efficient data taking technique is based on a Lock–In (LI) measurement as reported by Elzerman et al.[17] Again, we first pinch off one barrier completely by applying a sufficiently negative voltage to $TG_1$ (the barrier with more disorder is now completely closed). We then tune $TG_2$ to a desired tunnel rate. The CNT source and drain are both kept at ground potential and the SET charge detector is operated at a point of maximal charge sensitivity. Additional to the DC voltage, we apply a square pulse with equal up–down–times, $\tau$, to SG. We measure the SET response to this pulse train using a LI measurement at frequency $f = 1/2\tau$. We denote this response as the SET–LI signal. Schematically, the method is shown in Figure 3. The essential point is that electron tunneling into and out of the CNT–QD causes a dip in the SET–LI signal.

This depth of the dip depends strongly on the ratio of the pulse frequency, $f$, and the tunneling rate, $\Gamma$: If $f \ll \Gamma$, there is enough time for an electron to tunnel into the QD during the high phase of the pulse. In this case the dip has a 100% depth. Raising $f$ above $\Gamma$, electron tunneling becomes too slow to occur within a pulse cycle. As a result the dip depth will gradually decrease when increasing $f$ compared to $\Gamma$. Analytically, the dip depth is proportional to $1 - \pi^2 / (\Gamma^2 \tau^2 - \pi^2)$, under the assumption that rates for tunneling in and out are equal [17].

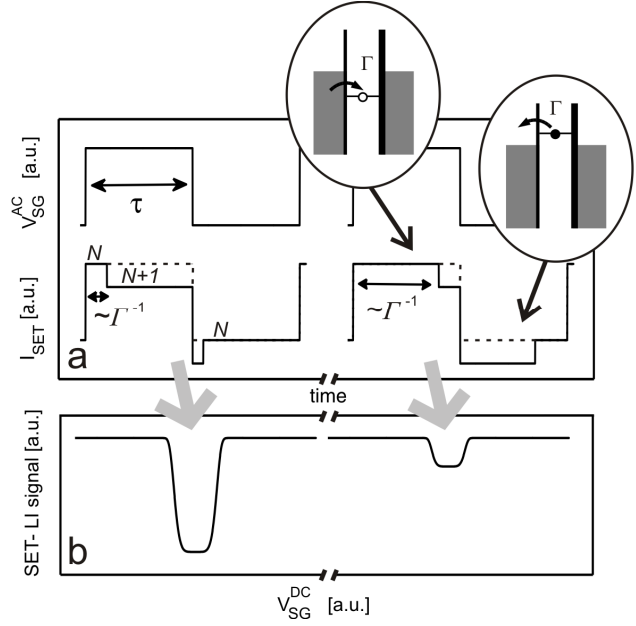

**Figure 3** Pulse spectroscopy scheme to determine the tunnel rate $\Gamma$ **a)** Square pulses with equal up– and down–times, $\tau$, are applied to the SG (top). Bottom: Schematic response of the SET–current to the pulse train. When the electron number in the CNT–QD remains constant (away from a charge transition), the SET current is only changed via the cross–coupling to the SG (dashed line). However, if we set the value of the SG voltage close to a charge transition of the CNT–QD, electron tunneling into and out of the QD will also affect the SET–current. During the high phase of the pulse an electron can tunnel into the QD on a timescale $\Gamma^{-1}$ (the tunnel rate through the open barrier) and leave it again during the low phase. This lowers (raises) the SET–current during the high (low) phase of the pulse (solid line). To illustrate why this method is sensitive to the tunnel rate, we plot the SET–current for a high tunnel rate ($\Gamma \gg f$, left) and a lower tunnel rate ($\Gamma \geq f$, right). The schematic energy diagrams show the electrochemical potential of the CNT–QD during the high (left) and the low (right) phase of the pulse. **b)** Expected SET Lock–In (LI) signal as function of the DC voltage applied to the SG. The SET signal is measured with a LI technique at frequency $f = 1/2\tau$. Without electron tunneling, the SET–LI signal stays constant. If electrons tunnel in and out the CNT–QD with a rate $\Gamma \gg f$, the SET–LI signal is lowered and shows a deep dip (left). If $\Gamma \geq f$ (right), the dip is less deep. And finally, if $\Gamma < f$, electron tunneling is too slow to occur within a pulse cycle and the dip will disappear completely. Therefore, measuring the SET–LI signal as a function of pulse frequency, $f$, allows us to determine the tunnel rate, $\Gamma$.

Figure 4a shows the SET–LI signal for different pulse frequencies. For low $f$ the effect of electron tunneling is large (pronounced dip) while the dip gradually disappears when raising $f$. From the fit in Figure 4c we obtain a tunnel rate $\Gamma = 1.4$ kHz. In order to set a desired value for the rate, we tune the voltage on $TG_2$. Indeed, Figure 4b shows that the size of the dip can be set to any value, each value corresponding to a particular rate. Again, this demonstrates the precise control over the tunnel rate in the interesting range of very long timescales.

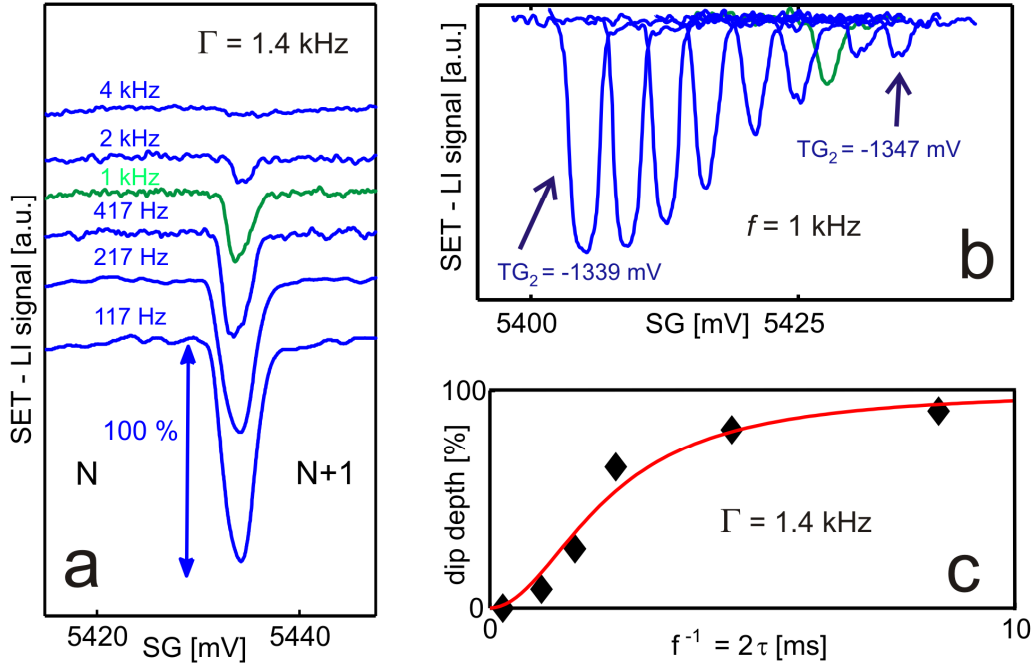

**Figure 4** Tuning and measuring the tunnel rate $\Gamma$ **a)** SET–LI signal for different pulse frequencies $f = 1/2\tau$ as a function of the DC–voltage applied to the SG. The SET is set to a configuration of highest sensitivity. For low pulse frequencies we observe a deep dip due to electron tunneling into the CNT–QD. At higher frequencies the dip gradually disappears. Curves are offset for clarity. **b)** SET–LI signal for a fixed pulse frequency $f = 1$ kHz for different voltages applied to $TG_2$. By lowering the voltage at $TG_2$ (from left to right), the dip becomes less deep. This shows that the tunnel rate into the CNT–QD changes from $\Gamma \gg 1$ kHz ($TG_2 = -1339$ mV) to $\Gamma \leq 1$ kHz ($TG_2 = -1347$) (note that when lowering the voltage on $TG_2$, the charge transition moves to a higher SG voltage due to the capacitive coupling of $TG_2$ to the CNT–QD). The green curves in a) and b) are taken at the same gate voltage and pulse frequency. We have used $TG_1 = -1800$ mV and BG = 5 V. **c)** Fit of the analytic expression for the dip depth in the LI signal to the measured data from a). The fit gives a tunnel rate $\Gamma = 1.4$ kHz. The 100% scale corresponds to the 100% arrow in a).

So far we have only considered tunneling into the ground state (GS) of our CNT–QD because we used relatively low pulse amplitudes. In the following, we investigate excited states (ES) of our closed

CNT–QD. We use the same pulse spectroscopy scheme as described above, but apply larger pulse amplitudes [17]. Now, we observe a shoulder–like feature in the dip of the SET–LI signal (Figure 5a). This feature is due to an ES of the CNT–QD: If the pulse amplitude exceeds the splitting between GS and ES, the effective tunneling rate into GS and ES is larger than the rate into the GS only. This results in a deeper dip in the SET–LI signal.

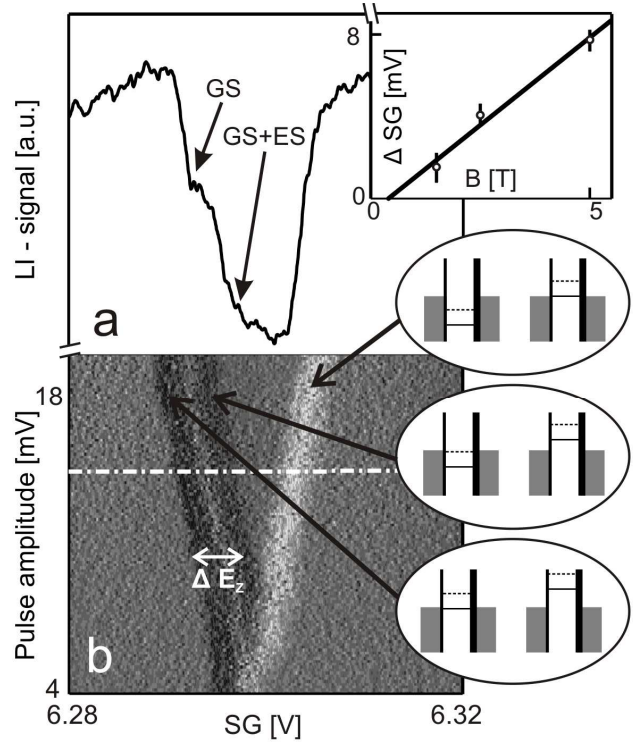

**Figure 5** Excited state spectroscopy **a)** SET–LI signal for a high pulse amplitude of 15 mV at $B$ = 2.5T. The gate settings are BG = 6.25 V, $TG_1$ = -2225 mV, $TG_2$ = -1870 mV and the pulse frequency is $f$ = 0.5 kHz. The dip now shows a shoulder like feature which is due to an increased tunnel rate into the CNT–QD when an excited state becomes accessible. Tunneling into both GS and ES state has a higher effective tunnel rate than tunneling into only the GS. Therefore, the dip becomes deeper when tunneling into the ES becomes energetically possible as well. **Inset:** splitting between GS and ES as function of magnetic field **b)** Derivative with respect to the sidegate voltage of the SET–LI signal as function of the DC voltage on the SG and pulse amplitude. Three lines are visible, whose meaning become clear when comparing to the single trace in a) and the schematic energy diagrams at the high and low phases of the pulse. The leftmost black line represents the onset of tunneling into the GS (lowest energy diagrams). The middle black line (which runs parallel to the first line) indicates where the dip becomes deeper because tunneling into the ES is possible as well (middle energy diagrams). The rightmost white line

shows the end of the dip (highest energy diagrams). The Zeeman splitting between GS and ES is indicated with the white arrow. The trace in a) is taken along the dashed–dotted line.

In Figure 5b we plot the derivative of the SET–LI signal as function of pulse amplitude and DC voltage on the SG. The excited state appears as a line, parallel to the onset of the dip. The distance between this line and the onset of the dip is the energy difference between GS and ES, converted to SG voltage by a factor $α$, the capacitance lever arm of the SG to the CNT–QD.

We identify the nature of the excited state by investigating its dependence on perpendicular magnetic field. We find that the splitting between GS and ES depends linearly on the magnetic field (Figure 5a, inset). Therefore, we conclude that GS and ES are spin–up and spin–down states of the same orbital level in the CNT–QD. Indeed, we expect that the Zeeman splitting at a magnetic field of a few Tesla is much smaller than orbital excitations. If we assume a length of ~500 nm for our CNT–QD, the orbital level spacing is on the order of meV [8, 9]. For the Zeeman splitting $E_Z = g\,μ_B\,B$, however, we expect $E_Z = 0.3$ meV at $B = 2.5\ T$ (assuming $g = 2$ [3, 18, 19]). We can use the energy of the Zeeman splitting to obtain the lever arm $α$ of the SG to the isolated CNT–QD and find $α = 0.065$.

In conclusion, we have investigated a CNT–QD that is nearly isolated from its leads. We have used an aluminum SET as a charge detector to read out the charge state of the isolated CNT–QD and measured the tunnel rate into the CNT–QD using both real–time charge sensing and pulse spectroscopy. We have found that it is possible to tune an individual tunnel barrier with high accuracy to very low tunnel rates, comparable to the expected spin relaxation and coherence times in CNTs. Finally, we measured the spin states of the nearly closed CNT–QD.

ACKNOWLEDGMENT We acknowledge technical assistance from R. Schouten, B. van der Enden and R. Roeleveld. We thank R. Hanson for discussions, C. M. Marcus and H. Churchill for allowing the use of their ALD – deposition equipment and M. Rinkiö for experimental help and discussion